\documentclass[aps,prc, groupedaddress,superscriptaddress,twocolumn,showpacs]{revtex4}
\usepackage{graphicx}
\usepackage{longtable}
\usepackage{array}
\begin {document}
\author {Yu.~M.~Gavriljuk}
\author {V~.N.~Gavrin}
\author {A.~M.~Gangapshev}
\author {V.~V.~Kazalov}
\author {V.~V.~Kuzminov}
\affiliation{Baksan Neutrino Observatory INR RAS, KBR, Russia}

\author {S.~I.~Panasenko}
\author {S.~S.~Ratkevich}
\affiliation{Karazin Kharkiv National University, Ukraine}

\date{\today}

\title{Comparative analysis of spectra of the background of the proportional \\
counter filled with Kr, enriched in $^{78}$Kr, and with Kr of natural content}

\begin{abstract}
The results of the experiment searching for 2K-capture with
large low-background proportional counter are presented. The
comparison of spectra of the background of the proportional
counter filled with Kr enriched in $^{78}$Kr (8400 hr) and with
natural Kr (3039 hr) is given.  A new limit on the half-life of
$^{78}$Kr with regard to 2K-capture,
T$_{1/2}\geq2.0\cdot10^{21}$ yrs ($95\%$ C.L.) has been
obtained.
\end{abstract}

\maketitle

 \section{Introduction}
The up to date experimental limit on the half-life of $^{78}$Kr
with regard to 2K(2$\nu$)-capture is
T$_{1/2}\geq1.5\cdot10^{21}$ yrs (90\% C.L.) [1,2]. Theoretical
calculations for this process based on different models predict
the following half-lives for $^{78}$Kr: $3.7\cdot10^{21}$ yrs
\cite{3}; $4.7\cdot10^{22}$ yrs \cite{4}; $7.9\cdot10^{23}$ yrs
\cite{5}. The two values, from \cite{4} and \cite{5}, have been
obtained by estimating the half-life of $^{78}$Kr with regard
to the total number of 2e(2$\nu$)-captures, taking into account
the portion of 2K(2$\nu$)-capture which constitutes 78.6\%.

Comparison of experimental and theoretical results shows that
sensitivity of measurement has reached the lower limit of
theoretical predictions. This work presents a direct
continuation of work \cite{1} and differs from it in collected
time of statistics which has increased from 159 hr to 8400 hr
for measurements with enriched krypton.

\section{The technique of the experiment}
The reaction $^{78}$Kr(2e$_k$,2$\nu$)$^{78}$Se yields an atom
of $^{78}$Se$^{**}$ with two vacancies on K-shell. The
technique of the search for this reaction is based on the
assumption that the energies of the characteristic photons and
probabilities of their emission for filling the double vacancy
coincide with the corresponding values for filling single
vacancies of K-shell in two atoms of Se$^{*}$   with single
ionization each. In such a case the total registered energy is
2K$ _{ab}$=25,3keV, where K$_{ab}$ is the binding energy of
K-shell's electron in the atom of Se(12.65 keV). Fluorescence
yield, due to the filling of the single vacancy on K-shell of
Se, is 0.596. The energies and relative intensities of the
characteristic lines of K-series are K$_{\alpha1}=11.22$ keV
(100\%), K$_{\alpha2}=11.18$ keV (52\%), K$_{\beta1}=12.49$ keV
(21\%), K$_{\beta2}=12.65$ keV (1\%) \cite{6}. Probabilities of
de-excitation with emission of  Auger electron ($e_a,e_a$)
only, or a single characteristic quantum and an Auger electron
(K,$e_a$), or two characteristic quanta and low energy Auger
electrons (K,K,$e_a$) are $\texttt{p}_1=0.163$,
$\texttt{p}_2=0.482$ and $\texttt{p}_3=0.355$, respectively. In
a gas, a characteristic quantum can pass a large enough
distance from its place of origin to the place of its
interaction. For instance, in krypton at a pressure of 4.35 atm
($\rho=0.0164$ g/cm$^{3}$) 10\% of the characteristic quanta
with energies of 11.2 keV and 12.5 keV are absorbed at a
distance of 1.83 and
\begin{figure}
\includegraphics[width=\linewidth]{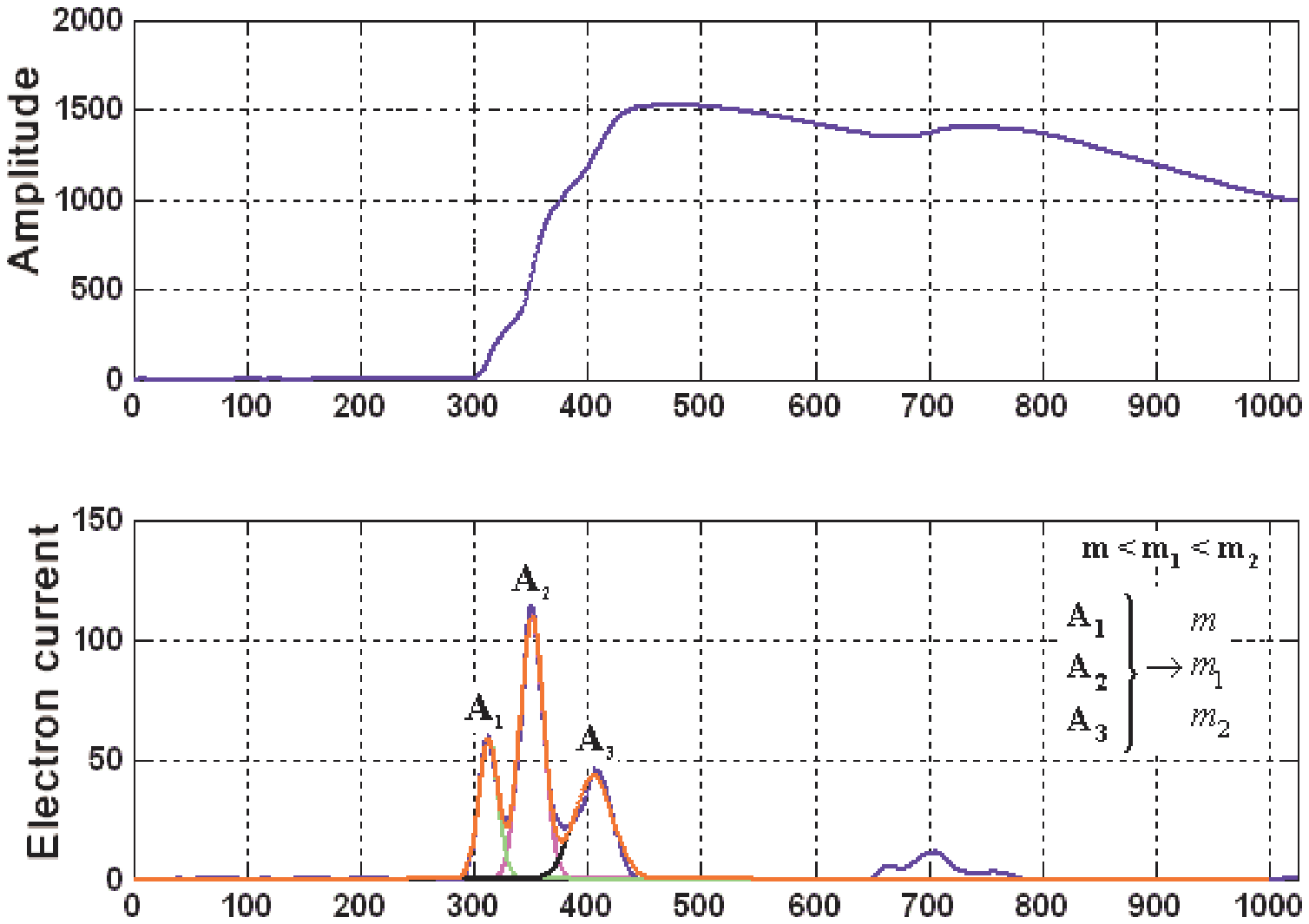}
 \caption{Example of the real three-point
pulse.}
\label{pic1}
 \end{figure}
2.42 mm, respectively (data on absorption coefficients were
taken from work \cite{7}). Runs of photoelectrons with the same
energies are equal to 0.37 mm and 0.44 mm, respectively. They
produce almost point-like energy release. In case of an event
with emission of two characteristic quanta and their subsequent
absorption in the working gas, the energy would be distributed
in three point-like regions. It is these events with a unique
set of characteristics that are the subject of the search
presented in this paper (fig.\ref{pic1})

To register the 2K-capture process the large proportional
counter (PC) with a body of copper of M1 type is used. The
cylindrical body has the following inner dimensions: diameter
of 140 mm, length of 710 mm. The anode wire made of gold-plated
tungsten is stretched inside the cylinder along its axis, its
diameter is 10 $\mu$m. The total volume of the counter is 10.4
l, the fiducial one is 9.16 l. PC is surrounded with
low-background shield of 8 cm boron polyethylene (BP) + 15 cm
Pb + 18 cm Cu. Pressure of krypton in the PC is 4.51 atm. In
basic measurements we use the sample of krypton with volume of
47.65 l enriched in $^{78}$Kr up to 99.81\% at the FSUE PA
"Electrochemical plant" of Zelenogorsk city. The set-up is
located in the separate room of the underground laboratories of
the Galium Germanium Neutrino Telescope of the Baksan
Underground Neutrino Observatory, INR RAS, at a depth of 4700
m. w. e.

\begin{figure}
\includegraphics[width=\linewidth]{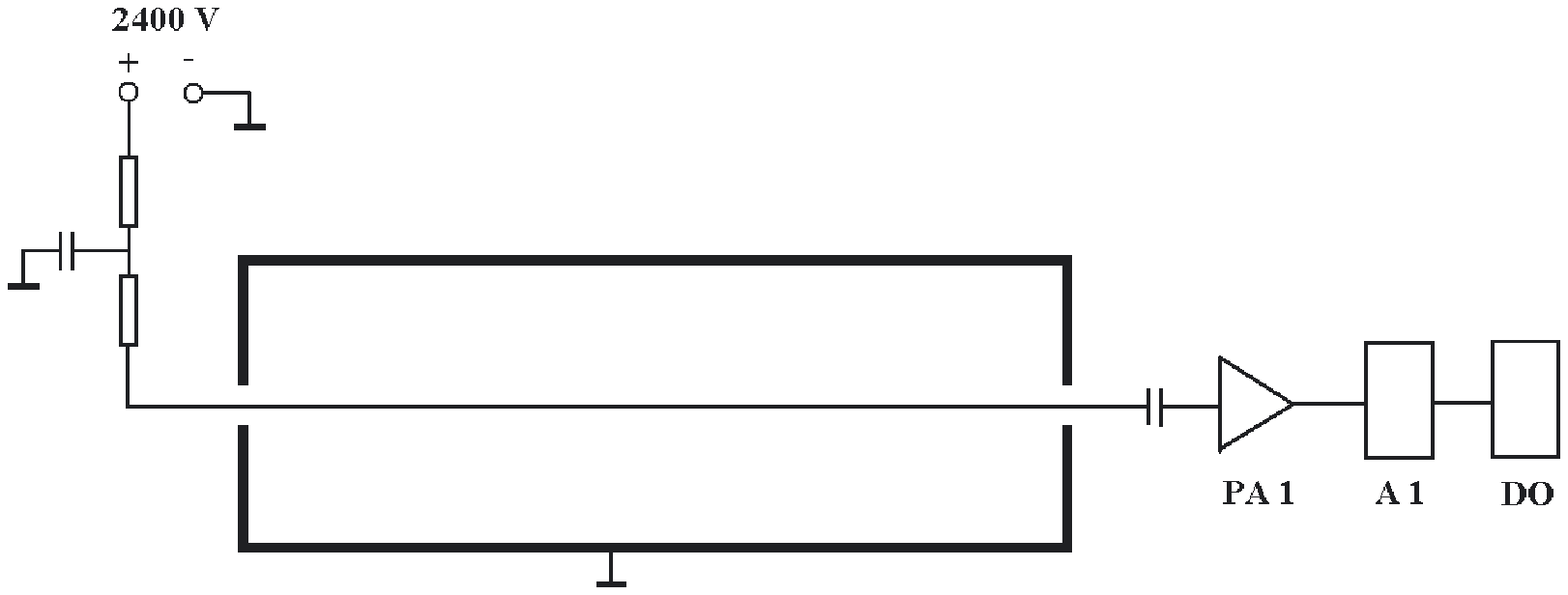}
 \caption{The electric scheme of installation. PA1
- preamplifiers, A1 - amplifiers,   DO - digital oscilloscope.}
\label{pic2}
 \end{figure}

\begin{figure}
\includegraphics[width=\linewidth]{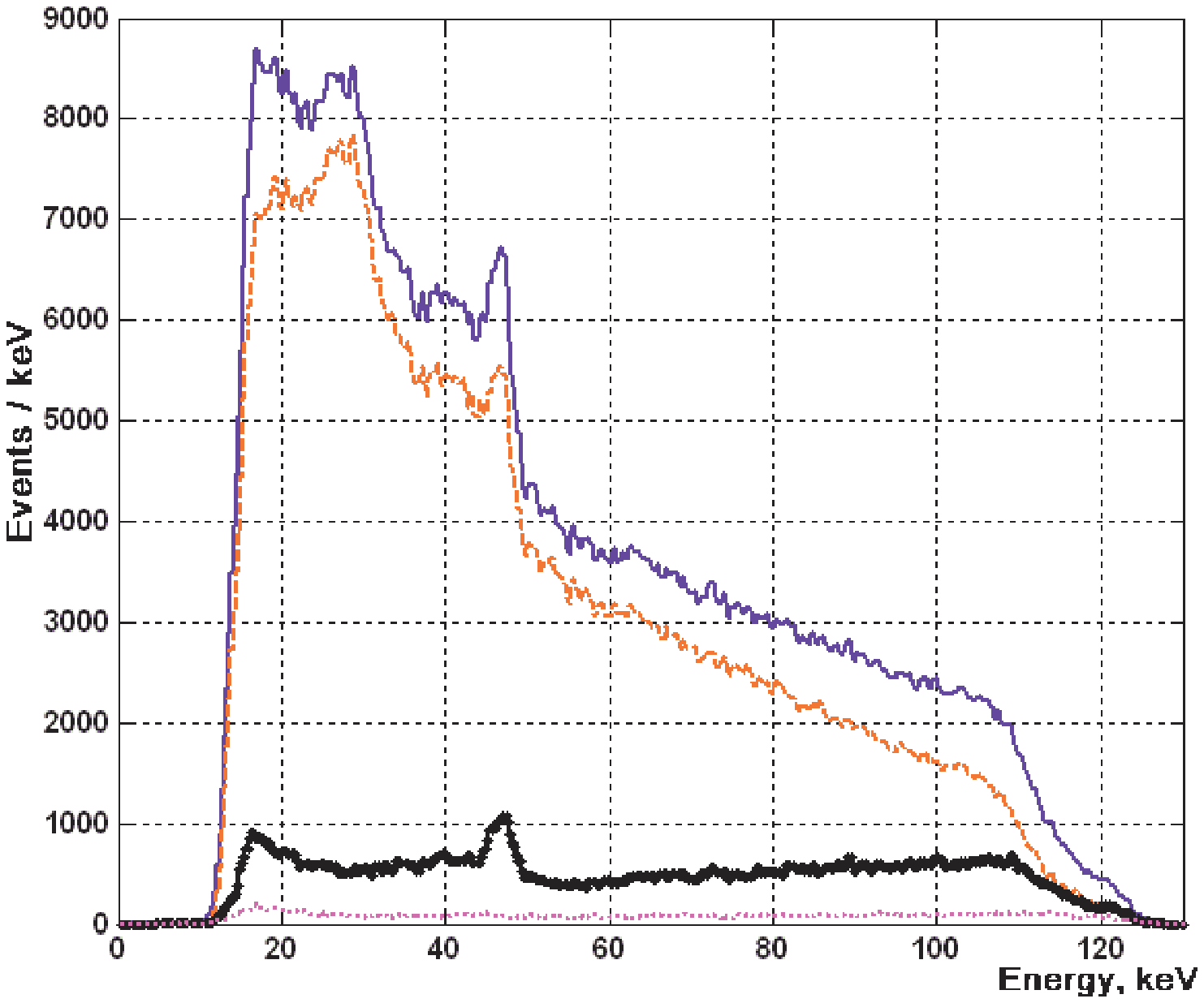}
 \caption{ Spectrum of PC's background for
krypton enriched in  $^{78}$Kr (8400 hr): solid curve - all the
events; dashed curve - single point-like events; dot-and-dash
line - two point-like events; dotted line - three point-like
events.}
\label{pic3}
 \end{figure}

The total electrical design of the apparatus  is presented in
fig.\ref{pic2}. High voltage (2400 V) is applied to the anode.
A signal passes via the high-voltage coupling capacitor to the
charge-sensitive preamplifier (PA). After amplification in the
additional amplifier (A) the pulses go to the input of the
digital oscilloscope (DA) LA-n20-12PCI incorporated into a
personal computer which writes down their shape digitized with
6.25 MHz frequency.

The registered pulses are then processed to determine the shape
of current pulses of the primary ionization electrons. Electron
current for a point-like event has the Gaussian shape. This
fact is used to determine the amount of the
similar-to-point-like components in each event by modeling it
with a set of Gauss functions (fig.\ref{pic1}). The amplitude
of each component (A) is determined by the area under the
Gaussian. Analysis of each component's amplitude is performed
for each event. The basic parameter of the analysis is the
ratio of amplitudes of the components in a current pulse
(fig.\ref{pic1}). For this purpose the amplitudes are  sized in
increasing order independently of their arrival time ({\it
m$_{i}$}). In addition to the main pulse there is a secondary
pulse in fig.\ref{pic1} which is due to photoelectrons knocked
out from the copper body of PC by photons generated in the
electron avalanches in the process of gas amplification. The
delay between the pulse and the after-pulse is determined by
the total drift time of electrons from the cathode to the
anode. It defines the time interval within which any single
event could be kept independently of the distribution of
primary ionization over the PC volume  .

In fig.\ref{pic3} are given spectra of PC's background
collected during 8400 hr with krypton enriched in $^{78}$Kr:
solid curve for all events, dashed curve for single point-like
events, dot-and-dash line for two point-like events and dotted
line for three point-like events. One can see the peak of total
absorption of $\gamma$-line for $^{210}$Pb at energy of 46.5
keV in the total spectrum (solid curve). Such a peak should be
mainly composed of single point-like events (photoeffect in
krypton with the de-excitation by Auger-electrons) and by two
point-like events (photoeffect with characteristic emission of
krypton).  Small part of events with absorption of the primary
(or characteristic) photon through its scattering on the outer
electrons with the subsequent followed by the absorption of the
secondary quantum would be three point-like.  Broad peaks,at
lower energies, are due to conversion electrons of the same
line (ce-L$_1$: Å=30.1 keV, 52\%; ce-Ì: Å=43,3 keV, 13,6\%
\cite{8}), coming from the surface copper layer. All the events
in these peaks should be single point-like. Conversion
electrons are accompanied with L-series characteristic quanta's
and low energy Auger-electrons (M-series characteristic quanta)
or with Auger-electrons from L-shell of $^{210}$Bi. Double
point-like event occur if the characteristic photon absorbed in
a working gas simultaneously with conversion electron coming
from the counter wall. A part of such events is small. This
quantitative picture is supported by the distribution of events
taken from different peaks of the spectra of corresponding
types. Comparison of the obtained ratio of areas of the peak of
46.5 keV for different components with theoretical calculations
allows one to determine the efficiency of the computer event
selection carried out for specified regions of ionization.

\begin{figure}
\includegraphics[width=\linewidth]{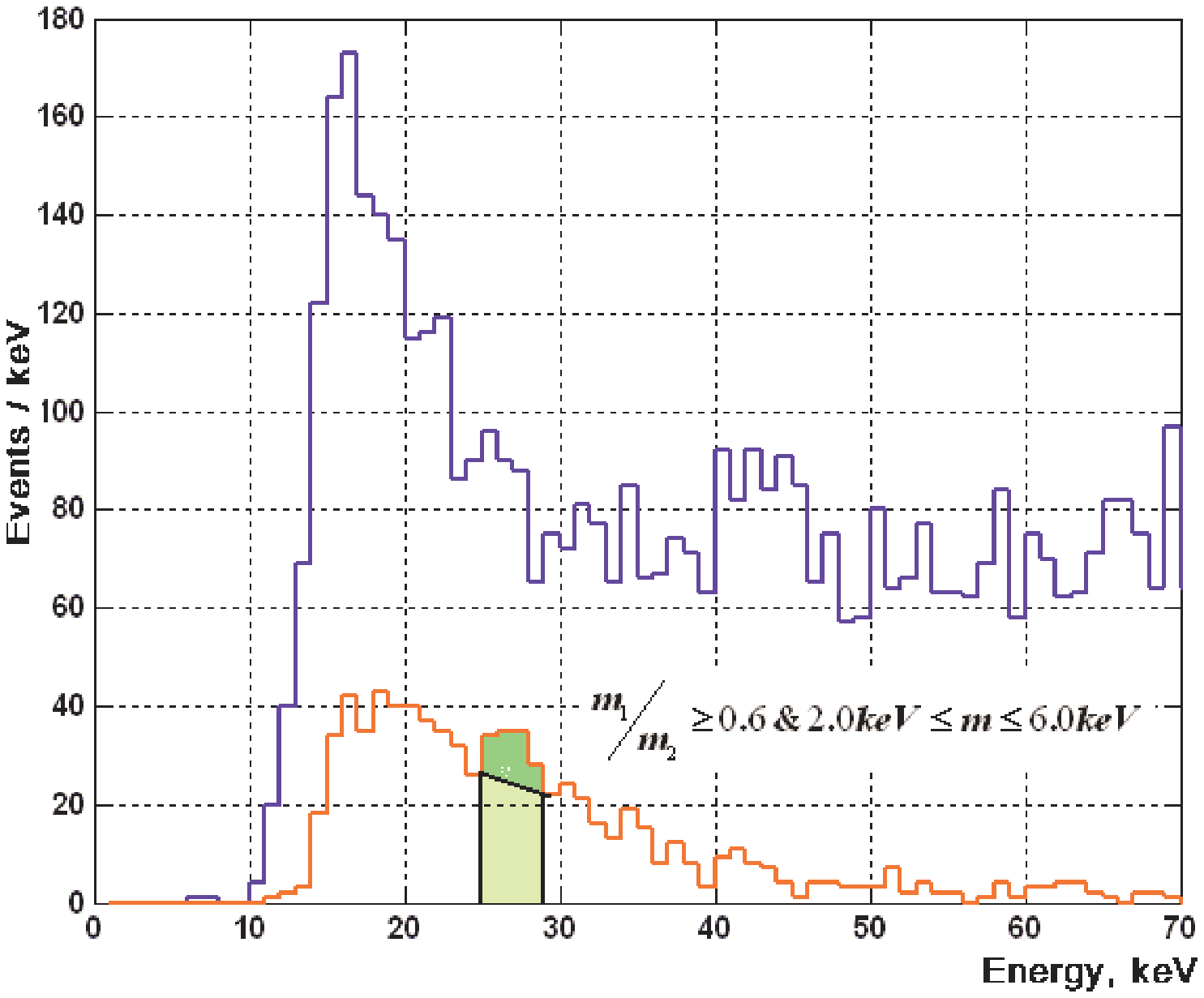}
 \caption{ Spectrum of three point-like
$^{78}$Kr events before and after the selection.}
\label{pic4}
\end{figure}
In fig.\ref{pic4} one can see the total spectrum of three
point-like events (the upper curve) and a spectrum of events
selected from the first one after application of the following
conditions on the components 2.0 keV $\geq m \geq$ 6.0 keV and
$m_1/m_2 \leq 0.6$ which include a set of characteristics
corresponding to the searched-for events (the lower curve). In
the second spectrum, one can see a peak at 25-30 keV which is
similar to the expected one for  the 2K-capture.

To clarify the nature of this peak an additional measurement
was carried out concerning the background of PC filled with
natural krypton (0.354\% $^{78}$Kr \cite{9}).  A radioactive
isotope $^{85}$Kr with volume activity of ~1000 Bq/l Kr is
present in natural krypton. It comes into the atmosphere mainly
through the processing of the debuggings of the fuel of atomic
stations. To eliminate this component from the PC's background
the sample of natural krypton was undergone an isotopic
purification by the way of centrifugation at the
Electrochemical Plant in Zelongorsk city.

\begin{figure}
\includegraphics[width=\linewidth]{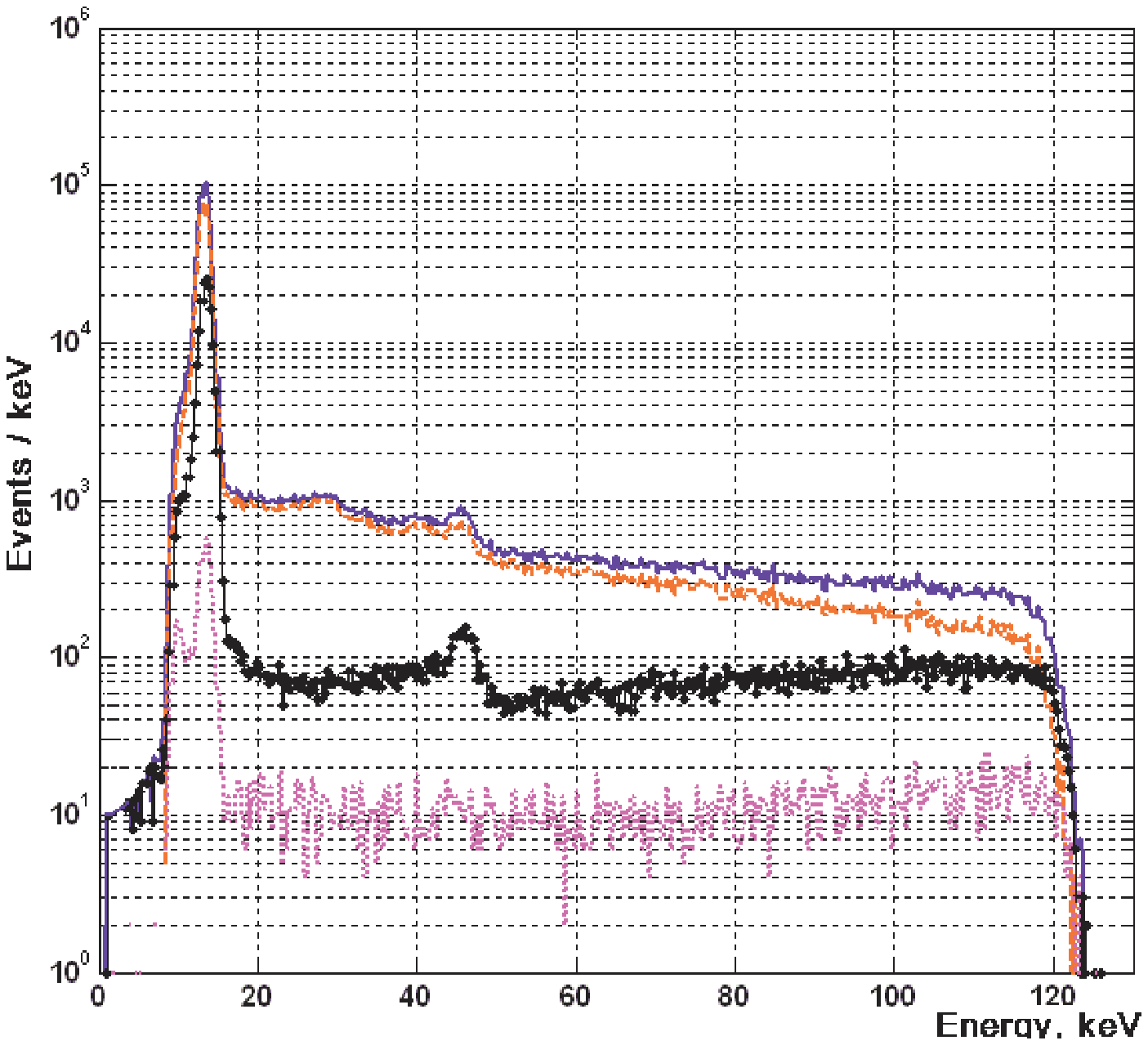}
\caption{Spectrum of the background of
PC filled with krypton of natural content during 3039 hours:
solid curve for all events; dashed curve for single point-like
events; dot-and-dash line for two point-like events; dotted
line for three point-like events.}
\label{pic5}
\end{figure}
In fig.\ref{pic5} are given spectra of the PC background
collected during 3039 hours with natural krypton at a pressure
of 4.51 at: solid curve for all events, dashed curve for single
point-like events, dot-and-dash line for two point-like events
and dotted line for three point-like events.   Atmospheric
krypton contains also cosmogenic radioactive isotope $^{81}$Kr
(Ò$_{1/2}=2.1\geq10^5$ yrs \cite{9}) with volume activity of
$~0.1$ min$^{-1}$l$^{-1}$ Kr [10,11]. It decays by capturing an
electron and gives $^{81}$Br (K-capture  - 87.5\% \cite{12}).
In case of K-capture the release of energy is equal to 13.5
keV. One can easily see this peak in fig.\ref{pic5}.

\begin{figure}
\includegraphics[width=\linewidth]{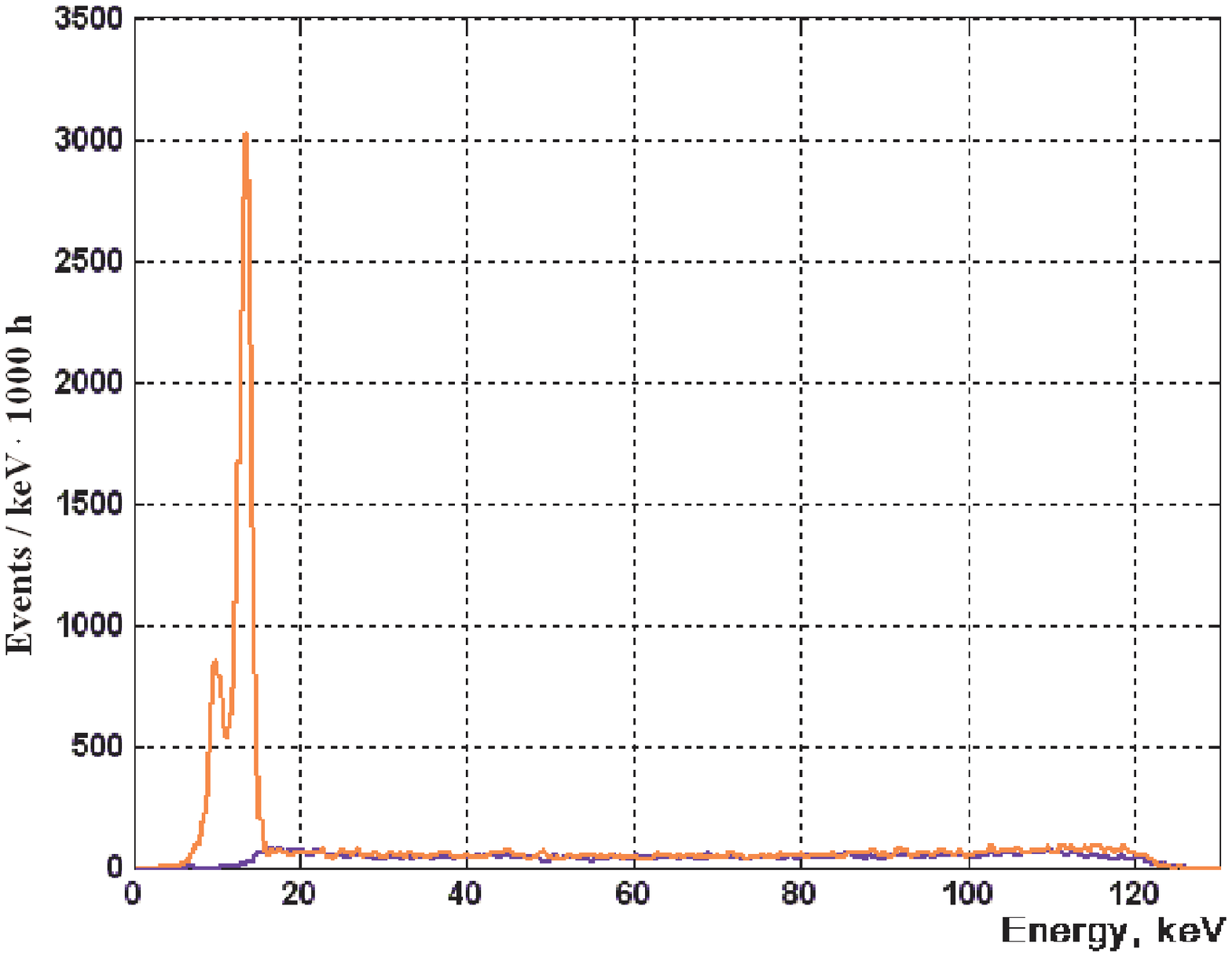}
\caption{Three point-like events for
natural krypton (solid curve) and for krypton enriched in
$^{78}$Kr (dashed curve).}
\label{pic6}
\end{figure}

Two spectra are presented for comparison in fig.\ref{pic6} for

\begin{figure}
\includegraphics[width=\linewidth]{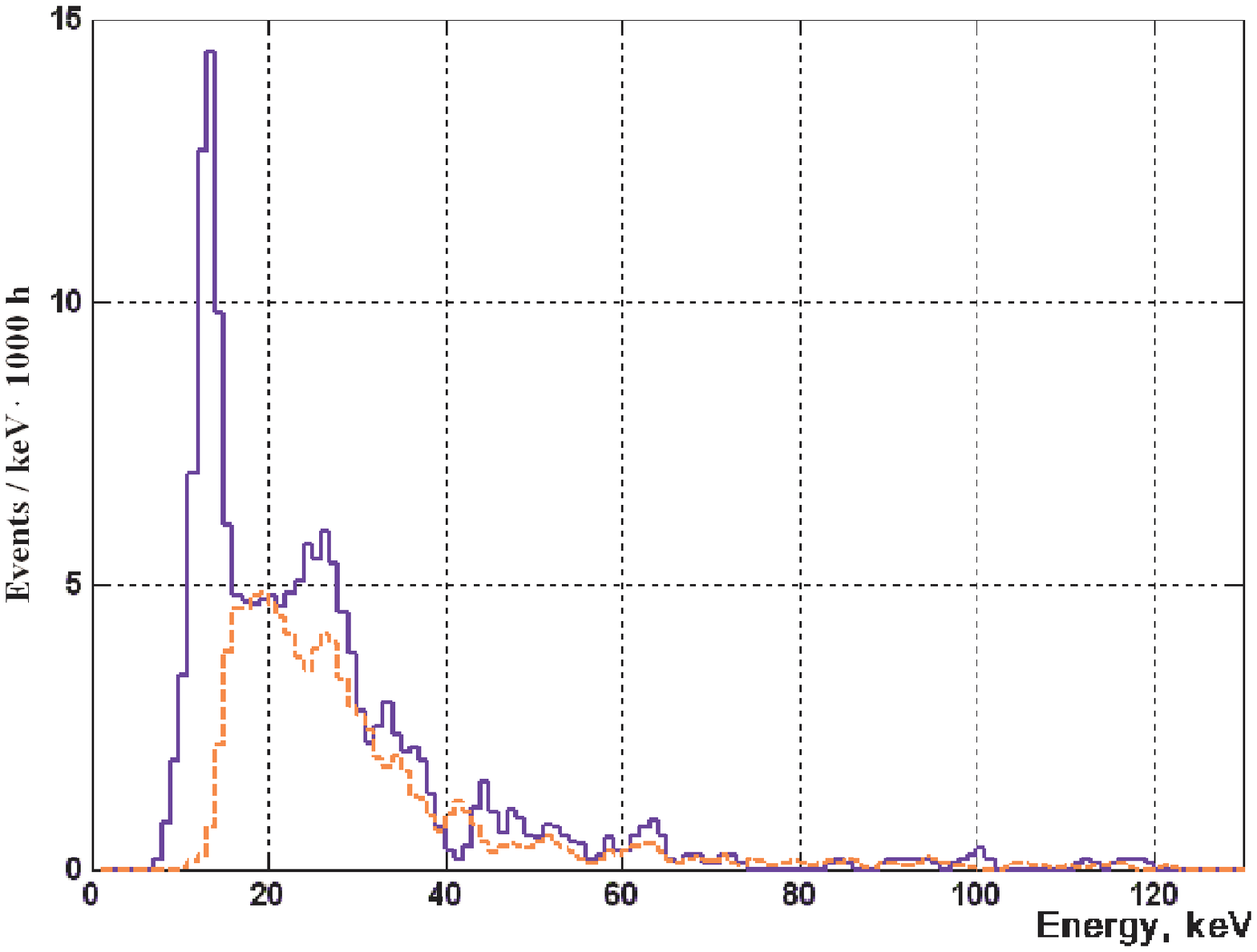}
\caption{Spectra for three point-like
events selected under mentioned conditions for natural krypton
(solid curve) and krypton enriched in $^{78}$Kr (dashed
curve).}
\label{pic7}
\end{figure}
three point-like events for krypton enriched in $^{78}$Kr
(dashed curve), and for krypton of natural content (solid
curve), scaled down to the same time. As is seen, they are in
good agreement. Slight difference seen at energies higher than
90 keV could possibly be explained by the presence of the
residual $^{85}$Kr in the sample of natural krypton. At high
enough energies the distribution of ionization along the track
of an electron in its projection onto the PC's radius could
simulate multiple-point events. Fig.\ref{pic7} shows the two
spectra for three point-like events selected under the
mentioned conditions for both samples. One can see that there
is a peak at 25-30 keV in natural krypton spectrum with the
intensity analogous to that of the peak in the spectrum for
krypton enriched in $^{78}$Kr. Presently we carry out thorough
investigation into the origin of this background peak. However,
it is possible to estimate the limit on the 2K-capture
contribution into the spectrum under study. In case we assign
all the observed effect to 2K-capture, the effect does not
exceed the area of the peak multiplied by a factor of two
standard deviations at a confidence level of 95\%.

Half-life limit was determined by:
\begin{eqnarray*}
\texttt{T}_{1/2}=\frac{\ln2 \times N\times \varepsilon_1 \times
\varepsilon_2 \times \texttt{p}_3 \times
(\texttt{p}_4+\texttt{p}_5)} {N_{\texttt{eff}}}
\end{eqnarray*}
where $\ln2=0.693$; $N=1.08\cdot10^{24}$ $^{78}$Kr atoms;
$\varepsilon_1=0.809$ is the efficiency of three point-like
events registration; $\varepsilon_2=0.43$ is the efficiency of
the three point-like events selection; $\texttt{p}_3=0.355$;
$\texttt{p}_4=0.713$; $\texttt{p}_5=0.213$;
$N_{\texttt{eff}}\leq46$ yr$^{-1}$, and was found to be
\begin{eqnarray*}
\texttt{T}_{1/2}(2K,2\nu)\geq
2\cdot10^{21}\,\,\texttt{yr}\,\,(95\% \,\,\texttt{C.L.}).
\end{eqnarray*}

The work has been carried out under the financial support of
the RFBR (grant no. 04-02-16037) and "Neutrino Physics" Program
of the Presidium of RAS.

\end{document}